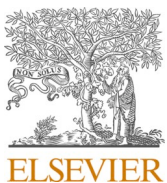

Contents lists available at ScienceDirect

# Surfaces and Interfaces

journal homepage: www.sciencedirect.com/journal/surfaces-and-interfaces

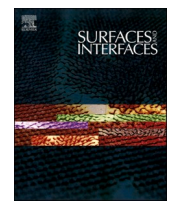

# A comprehensive study of the effect of thermally induced surface terminations on nanodiamonds electrical properties

Sofia Sturari [a,b], Veronica Varzi [a,b,c], Pietro Aprà [a,b,c,*], Adam Britel [a,b], Nour-Hanne Amine [a,b], Greta Andrini [b,d], Emilio Corte [a,b,c], Giulia Tomagra [c,e], Lorenzo Mino [c,f], Paolo Olivero [a,b,c], Federico Picollo [a,b,c]

[a] Physics Department, University of Torino, via P. Giuria 1, 10125 Torino, Italy
[b] National Institute of Nuclear Physics, Sect. Torino, via P. Giuria 1, 10125 Torino, Italy
[c] NIS Inter-Departmental Centre, via G. Quarello 15/a, 10135 Torino, Italy
[d] Department of Electronics and Telecommunications, Politecnico di Torino, Corso Duca degli Abruzzi 24, 10129 Torino, Italy
[e] Department of Drug and Science Technology, University of Torino Corso Raffaello 30, 10125 Torino, Italy
[f] Chemistry Department, University of Torino, via P. Giuria 7, 10125 Torino, Italy

## ARTICLE INFO

## ABSTRACT

Nanodiamonds (NDs) gained increasing attention in multiple research areas due to the possibility of tuning their physical and chemical features by functionalizing their surface. This has a crucial impact on their electrical properties, which are essential in applications such as the development of innovative sensors and in the biomedical field. The great interest in electrical conduction in NDs has driven the scientific community to its extensive investigation. The role of various functionalities has been considered and different conduction mechanisms have been proposed. In this work, we reported on a systematic study of the modification of the electrical properties of differently sized NDs, as a function of different thermal treatments in air, hydrogen or inert atmosphere. Samples were electrically characterized in controlled humidity conditions to consider the influence of water on conductivity. NDs electrical properties were interpreted in connection with their surface chemistry and structural features, probed with infrared and Raman spectroscopies. The presence of surface graphite, hydrogen terminations or water adsorbed on hydrophilic oxygen-containing functional groups rendered NDs more conductive. These moieties indeed enabled different conduction mechanisms, which were respectively graphite-mediated conduction, water-induced transfer doping and Grotthus mechanism in surface-adsorbed water. The effect of particles size on conductivity has also been evaluated and discussed. Our experimental findings shed light on the link between surface modifications and electrical properties of NDs, providing at the same time an interpretative key to understanding the effect of particles dimensions.

## 1. Introduction

Due to their appealing properties, including extreme hardness, high thermal conductivity, presence of photoluminescent defects [1–5] and excellent biocompatibility, nanodiamonds (NDs) are currently studied in a wide variety of research fields. These encompass tribology [6–8], catalysis [9–12], energy storage and conversion [13–15], photonics [16–18], quantum sensing [19–24] and biomedical applications [25–27]. These nanocrystals, whose size spans from few nm up to 300 nm, are typically obtained from detonation of carbon-containing explosives [28,29] or from milling of High-Pressure-High-Temperature

(HPHT) microdiamond powders [30]. Nonetheless, other production techniques can be exploited as well, as for instance Chemical Vapor Deposition (CVD) [31], laser ablation [32], high-energy ion irradiation of graphite [33] and synthesis from molecular precursors [34,35].

NDs typically exhibit a peculiar core-shell structure made up of a diamond core, a middle layer constituted by distorted $sp^3$ phases and amorphous carbon, and an outer shell formed by $sp^3$ and $sp^2$ carbon atoms with dangling bonds terminated by various moieties [36]. These surface functionalities are typically oxygen-containing species in as-synthesized NDs (*e.g.*, carbonyls, carboxyls, ethers, acid anhydrides and alcohol functions). However, surface groups can be easily tailored by exploiting wet chemistry approaches [37], exposure to microwave plasma [38], ball-milling processes [39], or high-temperature thermal treatments in controlled atmosphere [40].

As a result of their high surface-to-volume ratio, NDs final properties are heavily influenced by their surface structure. The study of the relationship between surface characteristics and properties thus represents an essential step in outlining NDs potential utilities and can also provide useful insights about their behavior in a broad range of applicative contexts.

In this framework, a relevant topic is represented by the connection between NDs surface modifications and electrical properties. It represents a complex issue, which has been previously explored from different viewpoints. Some of the previous works focused on the increase in electrical conductivity induced by surface graphitization [41]. For instance, Kuznetsov et al. investigated the resistivity of vacuum-annealed detonation NDs by means of four-point probe measurements. In that work, the particles resulted poorly conductive unless graphitic phases were formed, as it was the case of the NDs heated at temperatures larger than 1100 K [42]. Jackman's group instead performed impedance spectroscopy measurements at different temperatures on detonation NDs in the form of aggregates and individual particles [43,44]. They reported a loss of dielectric character occurring when exceeding a specific temperature threshold (namely, 350 °C and 400 °C for individual and aggregated NDs respectively), suggesting the onset of partial graphitization dynamics [43,44]. The link between graphite formation and NDs electrical properties was also studied by Čermák et al. They employed atomic force microscopy (AFM) to measure local electrical conductivity of HPHT NDs after thermal or plasma treatments, connecting it to a higher concentration of $sp^2$ bonds compared to the as-received NDs [45].

On the other hand, various papers explored the electrical conductivity of NDs upon their surface hydrogenation. It is well established that surface hydrogenation determines a negative surface affinity in diamond, thus inducing a substantial p-type surface electrical conductivity. The effect was explored in 2000 by Maier et al. for bulk diamond and it was interpreted on the basis of an electrochemical surface transfer doping model. This assumes that a thin water layer formed over the surface after atmosphere exposure acts as an electron acceptor system for hydrogenated diamond and thus promotes the formation of subsurface holes as charge carriers [46]. In more recent years, a mechanism similar to the one taking place in the bulk material was proposed for hydrogenated diamond nanoparticles, that also displayed a remarkable electrical conductivity. This was shown by Kondo et al., who reported that the resistivity of detonation NDs reduced from $10^7 \, \Omega \, \text{cm}$ to $10^5 \, \Omega \, \text{cm}$ upon thermal hydrogenation in the 600 °C - 900 °C temperature range [47]. They suggested that the observed increase of surface conductivity in hydrogenated NDs had to be attributed to the transfer doping model. The effectiveness of hydrogenation was indeed confirmed via Fourier transform infrared (FTIR) and X-ray photoelectron (XPS) spectroscopies. At the same time, the reduction of the $sp^2/sp^3$ carbon ratio at increasing temperature in the hydrogenation process allowed to rule out a possible role of $sp^2$ phases [47]. Welch et al. validated the findings by Kondo et al. with a characterization study of hydrogen-terminated NDs by means of impedance spectroscopy, carrying out the measurements both in air and in vacuum [48]. They found that for hydrogenated NDs the low impedance response was three orders-of-magnitude lower with respect to untreated NDs in air environment. The extremely high impedance registered when the hydrogenated samples were tested in vacuum corroborated the interpretation of hydrogen treatment inducing surface transfer doping in the nanoparticles. Indeed, the effect should not be seen in vacuum conditions, due to the removal of surface adsorbates [48]. The relatively high conductivity of hydrogenated NDs was also highlighted in the above-cited AFM analyses by Čermák et al. They proposed that this should exclusively stem from surface transfer doping mechanism, considering the efficacy of their hydrogenation treatment in establishing hydrogen terminations and the absence of amorphous or graphitic carbon on the NDs [45].

Besides graphite and hydrogen terminations, also the simple presence of adsorbed water on the surface has been indicated to affect NDs electrical properties. Surface adsorbed water indeed generates hydroxonium ions ($H_3O^+$) and thus allows electrical conduction to take place via charge transport due to Grotthus mechanism (proton hopping) [49,50]. This effect is based on a chain process, implying the transfer of a proton from a $H_3O^+$ ion to an adjacent water molecule, thus causing the formation of a new $H_3O^+$ and a further proton hop from the ion to another water molecule. The phenomenon is based on sequential breaking and reforming of covalent and hydrogen bonds and can be summarized by the following reaction: $H_2O + H_3O^+ \rightarrow H_3O^+ + H_2O$. The evidence that NDs conductivity is influenced by surface adsorbed water has been provided in two previous works [49,51]. In the former one, Piña-Salazar et al. investigated the electrical conductivity of untreated detonation nanodiamonds after exposure to humid air and after thermal treatment in vacuum. They showed that conductivity rose with increasing quantity of adsorbed water, whereas it underwent a two orders-of-magnitude drop upon heat treatment in vacuum, owing to the elimination of adsorbed water molecules [49]. In the latter work, Denisov et al. instead measured the conductivity of chlorinated and untreated detonation-synthesis NDs pellets by varying water vapor relative pressure ($p/p_s$). They revealed that electrical conductivity increased by five to six orders of magnitude in the $0\% < p/p_s < 98\%$ range, because of water adsorption. Denisov et al. also particularly stressed the impact of oxygen-containing surface functional groups on adsorbed water amount and electrical conduction. A larger concentration of oxygenated moieties was found to translate into more relevant water adsorption capacity and thus higher electrical conductivity [51]. By virtue of easily forming hydrogen bonds with water, these species are indeed highly hydrophilic and hence favor the adsorption of water on the surface of NDs, thus determining the enhancement of Grotthus mechanism effect on electrical conductivity. This was recently demonstrated by our research group in a study focused on the effects of reducing and oxidizing thermal processes on NDs surface water adsorption. In such work, we examined differently-sized milled and detonation NDs, by combining current-voltage characteristics measurements and various spectroscopic techniques [50]. We found that the untreated NDs and the ones subjected to thermal processes in air were generally less resistive than those heated in inert atmosphere. This was consistent with the higher hydrophilicity of the formers, resulting from more abundant oxygen-containing terminations on their surface [50].

Despite the above-listed body of previous studies, further investigations on the effect of surface treatment on NDs conductivity are required. Specifically, there is the lack of studies in which the effect of different surface modifications and NDs particle size on electrical properties is systematically correlated and examined. The present work was carried out exactly with the goal of filling this gap. Through a systematic approach, we studied how NDs electrical properties are affected by different thermal treatments in controlled atmosphere, correlating them with surface chemical and structural changes and with a specific conduction mechanism. To this aim, we performed NDs electrical characterization in different and carefully controlled relative humidity conditions, in conjunction with spectroscopic analysis. We performed diffuse reflectance infrared Fourier transform (DRIFT) and Raman spectroscopies, thus gaining deep insight on NDs surface chemistry and structure. Furthermore, this characterization was carried out on diamond powders with different primary particles median diameter to also assess the impact of particles dimension on electrical conduction. With this approach, electrical, chemical and structural characterizations could be combined, and it was possible to concurrently consider the existence of multiple conduction mechanisms.

## 2. Materials and methods

### 2.1. Nanodiamonds samples and thermal treatments

The majority of the experimental campaign was carried out on Micron+ NDs characterized by a 240 nm median diameter and produced by ElementSix™ (referred as "medium NDs / m-NDs" in the following). Other two diamond particles batches were examined to evaluate the effect of the size on electrical properties. In parallel with m-NDs, MSY NDs from Pureon with a median diameter of 55 nm ("small NDs/s-NDs" in the following) and Micron+ microdiamonds from ElementSix™ with 6 μm median diameter ("μDs" in the following) were analysed. All the diamond particles were obtained from the milling of HPHT type-Ib single crystals and were analysed after undergoing various thermal treatments in controlled atmosphere performed in a tubular furnace, following the sequence displayed in the two schemes of Fig. 1a (for m-NDs) and Fig. 1b (for s-NDs and μDs). The two diagrams report the abbreviations chosen to indicate the different thermal processes, as well as the labels used to name univocally each sample, which summarize all the processing steps carried out on a specific batch of NDs.

Annealing procedures were conducted in inert environment, *i.e.*, under nitrogen flow or in vacuum, for 2 h at temperatures of 800 °C or larger. Annealing treatments at 800 °C were adopted to standardize NDs surface, by eliminating surface functional groups, while graphitizing amorphous carbon components on the surface, without undermining the diamond material [40]. The processes were carried out on both pristine or previously treated NDs and they were indicated as '*Ann*' in the following. Annealing procedures at temperatures higher than 800 °C were instead employed to induce NDs graphitization, regardless of possible degradation of the diamond phase [52]. In detail, they were performed at 900 °C, 950 °C and 1000 °C and they have been called respectively '*HT-Ann-900*', '*HT-Ann-950*' and '*HT-Ann-1000*' in the following. Air oxidation processes were performed, after the '*Ann*' step, to selectively remove sp² carbon, thus purifying NDs from graphitic layers and concurrently promoting the formation of surface oxygen-containing functionalities [53]. Two different oxidation treatments were carried out. The first one, '*Ox*', was performed at 475 °C for 12 h and was exploited as an intermediate step for further processing, whereas the second oxidation, '*Mild Ox*', was conducted in less harsh conditions, *i.e.*, at 400 °C for 30 min. Eventually, in order to create C=H bonds on the surface, two distinct hydrogenation processes under hydrogen flow were performed. The one labelled as '*H₂-I*' was conducted at 900 °C for 3 h after the '*Ann*' process of the untreated particles, while the hydrogenation indicated as '*H₂-II*' was carried out at 800 °C for 2 h after '*Ox*' treatment.

### 2.2. Electrical measurements

Electrical measurements on NDs were performed in controlled-atmosphere conditions with varying relative humidity level, to correlate the effects of thermal treatments on NDs electrical properties with surface terminations and their interaction with environmental water vapor. The samples were loaded into a cylindrical cell that was suitably 3D-printed thanks to a stereolithography (SLA) Form 3B 3D-printer by using Formlabs Durable Resin. The cell had a depth d = $(2.85 \pm 0.05)$ mm and an inner diameter D = $(7.50 \pm 0.05)$ mm and it was closed with two aluminum electrodes. Its sealing and the contact of the electrodes with the powders were ensured by two springs, applying a constant and uniform pressure.

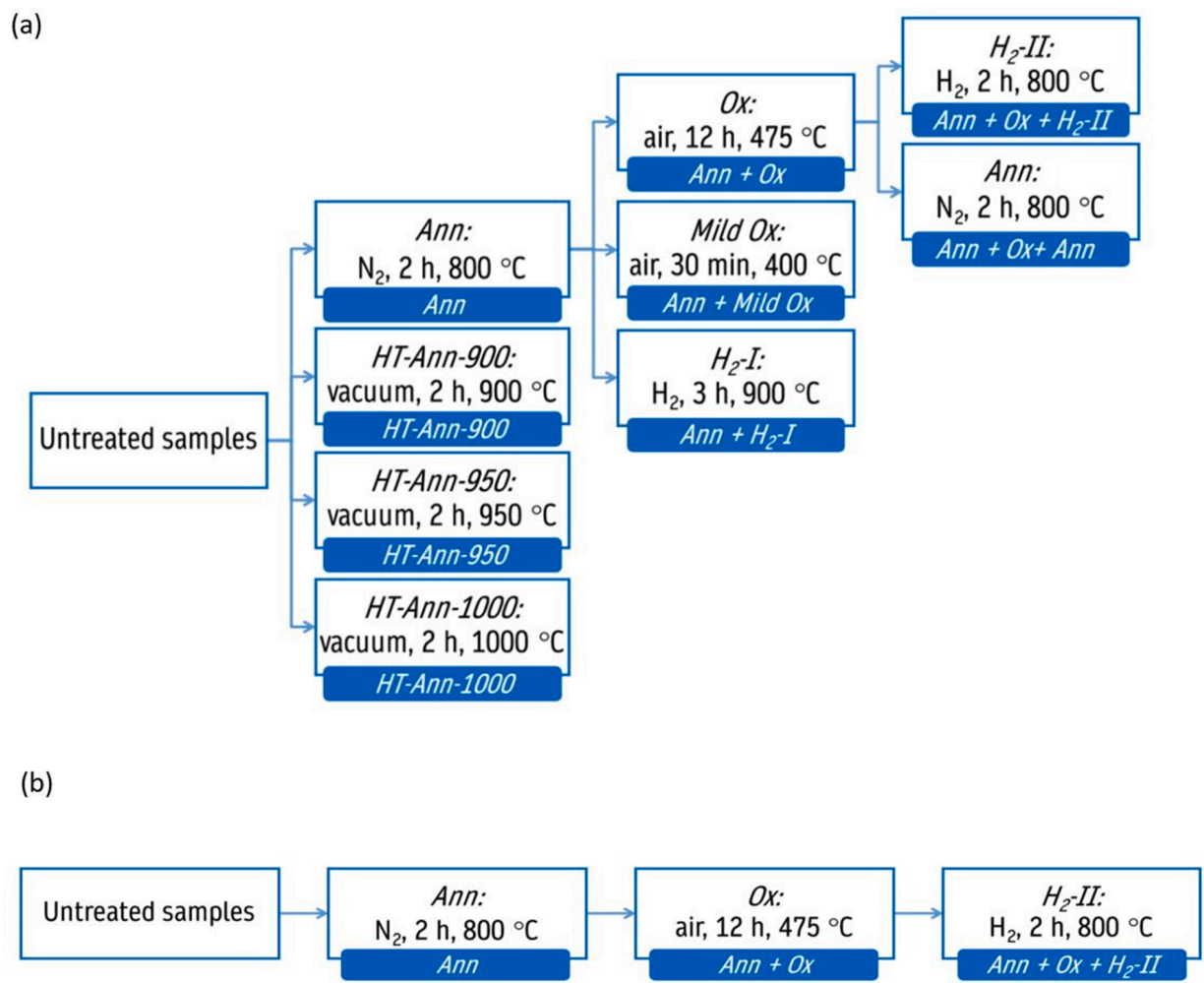

**Fig. 1.** (a) Thermal treatments carried out on m-NDs (240 nm median diameter). (b) Thermal treatments performed also on s-NDs (55 nm median diameter) and μDs (6 μm median diameter), besides m-NDs. The names of the processes were reported inside the white rectangles, whereas their sequence was indicated by the arrows. The names given to the samples were reported inside the blue rectangles: these summarize all the processing steps conducted on a given batch of NDs.

The NDs-filled cell was inserted into a probe station consisting of a vacuum chamber interfaced with a mechanical vacuum pump and connected to the apparatus used for electrical characterization [54]. The cell was left at a ~ 0.4 mbar pressure for 15 min to remove pristine adsorbed water on NDs and then the modulus of its impedance was measured. This quantity was considered to assess the electrical properties of the different samples and it has been indicated in the following as "NDs impedance" or |Z|.

A Hioki IM3536 LCR meter, operating in the 4 Hz – 8 MHz frequency range and displaying impedance modulus values up to 9.99999 GΩ, was used for recording NDs impedance values, setting a measuring voltage of 1 V and a frequency of 15 Hz.

After the above-described vacuum procedure, water vapor was introduced into the chamber, thanks to a valve connecting it to a water tank. A piezo-resistive Thyracont VD81 vacuum gauge was employed to measure the attained water pressure. The latter was then converted into a relative humidity value, by calculating the water saturation vapor pressure *via* the August-Roche-Magnus equation, reported below [55]:

$$p_s = 6.1094 \ e^{\frac{17.625 \cdot T}{T + 243.04}} \tag{1}$$

Where $p_s$ is water saturation vapor pressure expressed in mbar and T is the temperature in °C. The relative humidity of the chamber was kept fixed for 10 min, allowing NDs to adsorb environmental water until saturation and then the impedance was again measured. Afterwards, water vapor pressure was lowered and maintained constant for another 10 minutes. For each sample the described procedure was repeated to collect impedance values at progressively decreasing relative humidity levels, in the ~ 90% - 25% interval, in addition to the impedance value corresponding to 15 min vacuum exposure.

### 2.3. Raman spectroscopy

Raman spectroscopy was performed to structurally assess the NDs powders in terms of their graphitic/amorphous phase content. With this technique, possible structural modifications induced by thermal treatments were thus monitored. NDs were dispersed in isopropanol and deposited on a silicon wafer to collect Raman spectra, which were acquired with a Horiba Jobin Yvon HR800 Raman micro-spectrometer. This was equipped with a 600 lines mm$^{-1}$ diffraction grating, allowing a 3 cm$^{-1}$ spectral resolution, and a Peltier-cooled (-70 °C) CCD detector. The optical excitation was provided by a continuous Nd:YAG solid state 532 nm laser, focused with a 20 × objective. The employed magnification allowed probing a sample area of 10 × 10 μm$^2$ and a confocal depth of ~ 3 μm. The laser power intensity incident on the sample was regulated to 1.69 mW by inserting a filter along the optical path.

### 2.4. Diffuse reflectance infrared Fourier transform (DRIFT) spectroscopy

DRIFT spectroscopy was carried out to investigate NDs surface chemistry. DRIFT spectra were collected in diffuse reflectance mode using a Bruker Vector 22 FTIR spectrometer equipped with a mercury-cadmium-telluride detector. Each spectrum was acquired in dry air ambient conditions (*i.e.*, relative humidity ~ 10%), by averaging 128 acquisitions at 2 cm$^{-1}$ spectral resolution. The reflectance values were successively converted in pseudo-absorbance values (*i.e.*, A = - log R, where R is the measured reflectance).

## 3. Results and discussion

### 3.1. Characterization of m-NDs (240 nm median diameter)

The results of electrical characterization at variable relative humidity (indicated also as "RH" in the following) of the m-NDs subjected to the thermal treatments shown in Fig. 1a were presented in Fig. 2a, 3a and 4a and summarized in Table 1.

Fig. 2a displays the results of electrical measurements from the m-NDs labelled as '*Ann*', '*Ann + Ox*', '*Ann + Mild Ox*' and '*Ann + Ox + Ann*'. Raman spectroscopy was essential in revealing that no graphitization was induced on NDs by thermal treatments and allowed to *a priori* rule out any role of graphitic phases on electrical properties. The corresponding Raman spectra were reported in Fig. S1 of the ESI. From the graph in Fig. 2a it is possible to note that the impedance of the NDs, which exceeded the upper limit of the instrumental range when tested after keeping the samples in vacuum, decreased at increasing relative humidity values. This experimental evidence highlights a remarkable influence of adsorbed water on particles electrical properties and seems to suggest that such properties are mainly determined by proton hopping occurring in water. Our experimental data are indeed coherent with prior findings concerning non-graphitized NDs. These indicated an increase in conductivity with increasing amount of adsorbed water on NDs surface, due to an increasingly pronounced effect of the Grotthuss mechanism [49,50].

For the '*Ann*'-NDs, |Z| decreased from (6700 ± 1500) MΩ at the lowest relative humidity level to (620 ± 140) MΩ at the highest RH value. For the '*Ann + Ox*'-NDs, the impedance dependence on hydration conditions became even more prominent: the impedance variation was of two orders-of-magnitude, namely from (760 ± 170) MΩ to (3.2 ± 0.7) MΩ, over the RH interval under examination. Notably, impedances were relevantly lower in this case with respect to the previous one, the difference being of about one order of magnitude and increasing up to two orders of magnitude at the highest RH value. The '*Ann + Mild Ox*'-NDs also had a |Z| varying over two orders of magnitude, namely from (9000 ± 2000) MΩ to (27 ± 6) MΩ over the RH interval under examination. However, the impedance values of the '*Ann + Mild Ox*'-NDs were systematically larger than the ones of the '*Ann + Ox*'-samples by about one order of magnitude and were closer to the values measured for '*Ann*'-NDs when RH was reduced below 60%. '*Ann + Ox + Ann*'-NDs displayed impedance values that were close to those of '*Ann*'-NDs, as well as the same one-order-of-magnitude |Z| variation over the RH interval under examination.

An insight into the correlation between electrical properties, surface chemistry and adsorbed water for this batch of processed NDs was provided by their DRIFT spectra, which were presented in Fig. 2b. The DRIFT spectrum of '*Ann*'-NDs exhibits relevant signals in the 2990 cm$^{-1}$ - 2800 cm$^{-1}$ range (labelled as "$\nu_{C-H}$"), which was attributed to asymmetric and symmetric stretching of C-H bonds. The nearly complete absence of features in the 3650 cm$^{-1}$ - 3000 cm$^{-1}$ range (labelled as "$\nu_{O-H}$") was attributed to the lack of O-H stretching modes of both surface adsorbed water and surface functional groups [56]. In the spectral range associated to C≡O stretching (labelled as "$\nu_{C=O}$"), only a minor band at 1705 cm$^{-1}$ is visible. The dominant species after the '*Ann*' step are thus C-H moieties, whereas oxygen-containing groups are substantially absent. This implies a low affinity of '*Ann*'-NDs towards water, whose presence was indeed not detectable. Substantial spectral modifications were observed from '*Ann + Ox*'-NDs: their spectrum exhibits a broad band in the $\nu_{O-H}$ spectral range and extremely weak signals in the $\nu_{C-H}$ one. Moreover, in the $\nu_{C=O}$ spectral range it is possible to note a broad and complex band at 1800 cm$^{-1}$, that was ascribed to the C≡O stretching in different surface moieties, including carboxylic acids, esters, lactones, acid anhydrides, etc. [57–59]. The same surface oxygenated groups are also responsible for the appearance of new spectral features in the

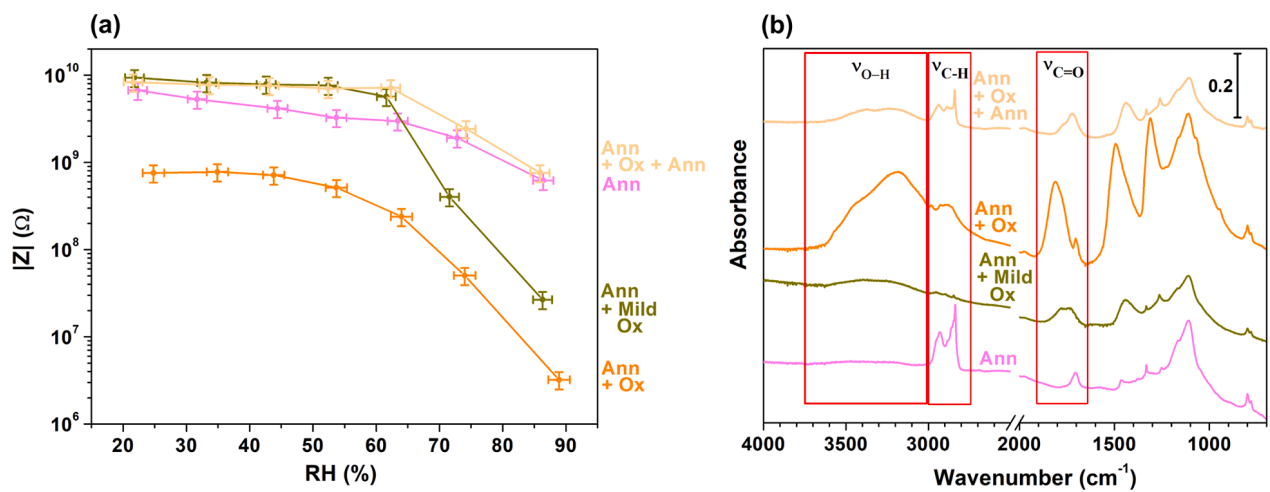

**Fig. 2.** Characterization of the '*Ann*'-m-NDs, '*Ann + Ox*'-m-NDs, '*Ann + Mild Ox*'-m-NDs and '*Ann + Ox + Ann*'-m-NDs (240 nm median diameter). (a) Electrical impedance of the samples at different relative humidity (RH) conditions. (b) Samples DRIFT spectra. The regions of O-H stretching ($\nu_{O-H}$), C=H stretching ($\nu_{C-H}$) and C=O stretching ($\nu_{C=O}$) were marked with red rectangles.

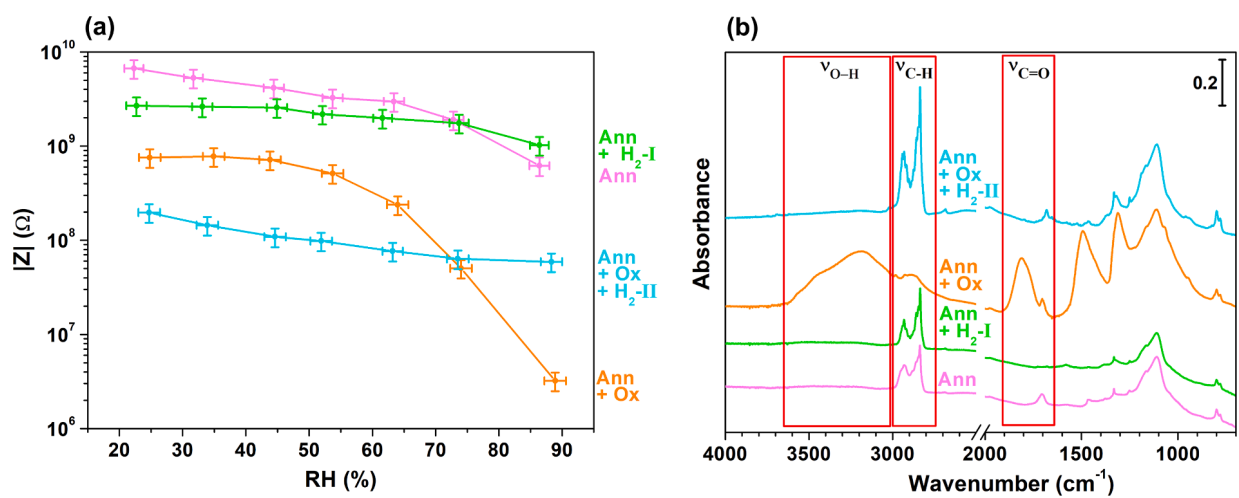

**Fig. 3.** Characterization of the '*Ann + H₂-I*'-m-NDs and '*Ann + Ox + H₂-II*'-m-NDs (240 nm median diameter). The data concerning the '*Ann*'-m-NDs and '*Ann + Ox*'-m-NDs have been also shown for sake of comparison. (a) Electrical impedance of the samples at different relative humidity (RH) conditions. (b) Samples DRIFT spectra. The regions of O=H stretching ($\nu_{O-H}$), C=H stretching ($\nu_{C-H}$) and C=O stretching ($\nu_{C=O}$) were marked with red rectangles.

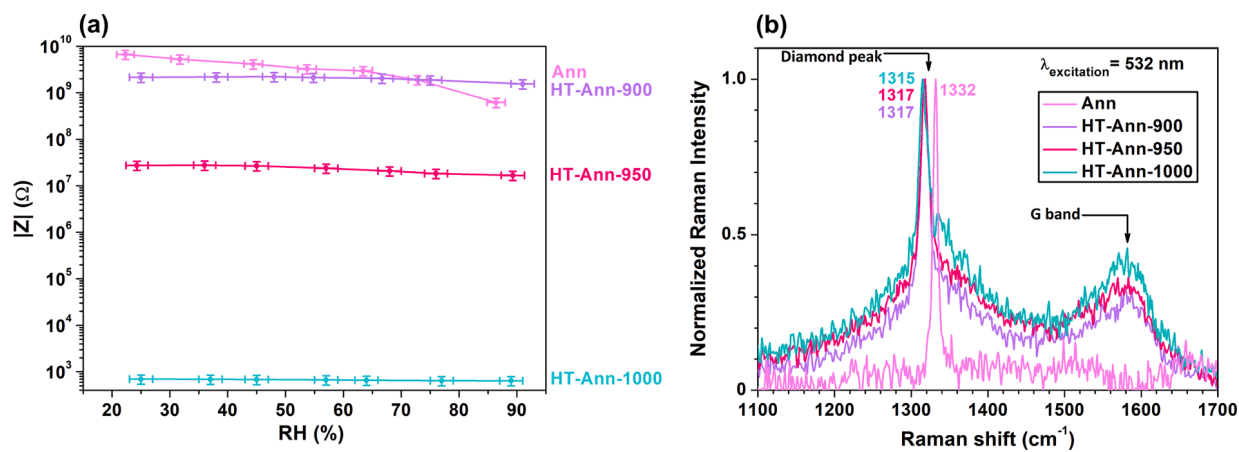

**Fig. 4.** Characterization of the '*HT-Ann-900*'-m-NDs, '*HT-Ann-950*'-m-NDs and '*HT-Ann-1000*'-m-NDs (240 nm median diameter). The data concerning the '*Ann*'-m-NDs have been also shown for sake of comparison. (a) Electrical impedance of the samples at different relative humidity (RH) conditions. (b) Samples Raman spectra. Diamond peak and graphite G band were highlighted by arrows. The position of the diamond peak was also reported.

**Table 1**

Impedance values measured from m-NDs (240 nm median diameter) at relative humidity ~ 25% (indicated as |Z|$_{25}$) and relative humidity ~ 90% (indicated as | Z|$_{90}$). The ratio between |Z|$_{25}$ and |Z|$_{90}$ was also reported.

| m-NDs sample | |Z|$_{25}$ (MΩ) | |Z|$_{90}$ (MΩ) | |Z|$_{25}$ / |Z|$_{90}$ |
|---|---|---|---|
| *Ann* (800 °C) | 6700 ± 1500 | 620 ± 140 | 11 |
| *Ann + Ox* | 760 ± 170 | 3.2 ± 0.7 | 238 |
| *Ann + Mild Ox* | 9000 ± 2000 | 27 ± 6 | 333 |
| *Ann + Ox + Ann* | 8300 ± 1900 | 760 ± 170 | 11 |
| *Ann + H₂-I* | 2700 ± 600 | 1000 ± 200 | 3 |
| *Ann + Ox + H₂-II* | 200 ± 40 | 59 ± 13 | 3 |
| *HT-Ann-900* | 2200 ± 500 | 1600 ± 300 | 1 |
| *HT-Ann-950* | 28 ± 6 | 17 ± 4 | 2 |
| *HT-Ann-1000* | 0.00070 ± 0.00016 | 0.00064 ± 0.00014 | 1 |

1500 cm$^{-1}$ - 1300 cm$^{-1}$ spectral region, typical of ketones and epoxides vibrations [57,58]. The previously commented spectral bands indicate that the '*Ox*' step led to the formation of oxygen-containing functionalities, thus producing increased surface hydrophilicity. The spectrum from '*Ann + Mild Ox*'-NDs is similar to that of '*Ann + Ox*'-NDs, being characterized by the suppression of C-H bands, the appearance of the $\nu_{C=O}$ and O-H stretching bands. These features indicate that also the less aggressive oxidation process produced oxygenated moieties and increased the water-affinity of NDs surface. Nonetheless, by comparing the spectrum from '*Ann + Mild Ox*'-NDs with the one from '*Ann + Ox*'-NDs, it is possible to notice that the O-H stretching and the $\nu_{C=O}$ bands, as well as the spectral features at the low wavenumbers, are much less intense in '*Ann + Mild Ox*'-NDs. These observations reflect considerable differences with varying oxidation conditions. The '*Mild Ox*' seems to cause to a restricted extent the formation of oxygenated groups and to not result in a highly hydrophilic surface, translating into a limited amount of adsorbed water on '*Ann +Mild Ox*'-NDs surface. On the other hand, in the spectrum of '*Ann + Ox + Ann*'-NDs, the disappearance of the $\nu_{C=O}$ band can be noted. Such disappearance is accompanied by a concomitant reduction of the O-H band, a slight enhancement of the bands between 2990 cm$^{-1}$ and 2800 cm$^{-1}$ and partial vanishing of the signals at low wavenumbers. The '*Ann + Ox + Ann*' treatment hence caused the eradication of oxygen-containing groups and the formation of C-H moieties, thus re-conferring a strong hydrophobicity to the NDs.

The smaller impedance values and the stronger |Z| dependence from relative humidity observed for '*Ann + Ox*'-NDs were thus explained by the presence of oxygenated moieties on the surface (such as hydroxyls, carboxylic acids, esters, etc.) introduced by the '*Ox*' step [50,57]. This determined a higher surface hydrophilicity and, in turn, an increased water adsorption capacity, consequently enhancing Grotthus mechanism and therefore electrical conduction. The weaker conductivity and impedance dependence from RH registered for '*Ann*'-NDs and '*Ann + Ox + Ann*'-NDs instead resulted from a reduced adsorption of water, which translated into a negligible effect of Grotthus mechanism on electrical conductivity and was connected with surface hydrophobicity. Such hydrophobicity was in turn produced by the '*Ann*' step, which eradicated oxygenated groups instead of forming them. Finally, the higher impedance values recorded for '*Ann + Mild Ox*'-NDs with respect to '*Ann + Ox*'-NDs originated from a more limited effect of Grotthus mechanism, caused by the lower amount of adsorbed water, in turn dictated by the lower surface density of oxygenated species. The smaller amount of oxygen-containing surface groups was due to the adopted oxidation conditions, which were much milder for the '*Ann + Mild Ox*'-NDs.

Contrarily to the samples discussed so far, the hydrogenated m-NDs displayed an impedance trend that was barely dependent on relative humidity conditions, as it can be seen from Fig. 3a. Fig. 3a reports the outcomes of electrical measurements from the '*Ann + H₂-I*'-NDs and '*Ann + Ox + H₂-II*'-NDs, as well as the ones from the '*Ann*'-NDs and '*Ann + Ox*'-NDs, for sake of comparison. However, in analogy with the NDs that yielded the data reported in Fig. 2, also these NDs were not enough conductive to be successfully tested by the measuring apparatus upon vacuum treatment. The influence of surface graphite on electrical behavior could be disregarded, since no graphite G-band was observable in hydrogenated NDs Raman spectra, as reported in Fig. S1 of the SI.

The impedance of '*Ann+ H₂-I*'-NDs was (2700 ± 600) MΩ at RH ~ 25% and reached (1000 ± 200) MΩ at RH ~ 90%, while the same parameter for '*Ann + Ox + H₂-II*'-NDs decreased from (200 ± 40) MΩ (as measured at the lowest RH value) to (59 ± 13) MΩ (as measured at the highest RH value). The dependence of the impedance from humidity was thus limited to only a three-fold factor for the hydrogenated samples. From the ten-fold impedance variation measured between the '*Ann + H₂-I*'-NDs and '*Ann+ Ox + H₂-II*'-NDs it is possible to appreciate that the latter ones were less insulating than the former. It is also worth comparing '*Ann + H₂-I*'-NDs and '*Ann+ Ox + H₂-II*'-NDs with '*Ann*'-NDs and '*Ann + Ox*'-NDs. It can be noted that for relative humidity less than 60% the '*Ann + H₂-I*'-NDs became more conductive than the '*Ann*'-NDs. Furthermore, also the '*Ann + Ox + H₂-II*'-NDs increased their conductivity value, compared to the '*Ann + Ox*'-NDs. For humidity levels around 25% and 35%, the impedance values of '*Ann+ H₂-I*'-NDs displayed a two-fold decrease with respect to '*Ann*'-NDs, whereas the impedance values of '*Ann+ Ox + H₂-II*'-NDs were lower by one order of magnitude with respect to '*Ann + Ox*'-NDs.

The large impedance values registered upon retaining hydrogenated samples in vacuum conditions and the negligible impedance variation with changing RH conditions can be rationalized by taking some assumptions on the electrical conduction mechanism and taking coherent considerations about surface chemistry of the processed samples. Since hydrogenation forms H terminations and typically causes etching of non-diamond carbon and the concurrent elimination of oxygenated groups [60], it should confer a hydrophobic character to NDs surface [61]. We could therefore assume that very limited water adsorption on such hydrophobic surfaces occurred, and hence that the Grotthus mechanism conduction was not relevant in this case. However, the minimal amount of surface adsorbed water could be sufficient to trigger surface transfer doping mechanism, through the formation of a subsurface hole accumulation layer. On the other hand, the effect of hole conduction in '*Ann + H₂-I*'-NDs and '*Ann + Ox + H₂-II*'-NDs could be evident only when relative humidity was reduced because of the relatively strong impact of proton hopping mechanism on the impedance values of '*Ann*'-NDs and '*Ann + Ox*'-NDs. The higher conductivity of '*Ann + Ox + H₂-II*'-NDs in comparison to '*Ann + H₂-I*'-NDs could be attributed to a higher efficiency of the hydrogenation process when carried out after the '*Ox*' step.

This interpretation of the electrical conduction mechanism and water adsorption in hydrogenated NDs was corroborated by the analysis of the DRIFT spectra collected from the '*Ann + H₂-I*'-NDs and '*Ann+ Ox + H₂-II*'-NDs, reported in Fig. 3b. In that same figure, also the spectra collected from the '*Ann*'-NDs and '*Ann + Ox*'-NDs were presented for sake of reference. In the DRIFT spectrum collected from the '*Ann + Ox + H₂-II*'-NDs it is possible to observe a significant enhancement of the bands linked to C-H stretching with respect to the spectrum collected from the '*Ann + Ox*'-NDs. Moreover, the suppression of a part of the spectral features in the range below 1500 cm$^{-1}$ is also observable. The enhancement of features attributed to C-H stretching is less marked but still clear in the spectrum collected from '*Ann + H₂-I*'-NDs, as compared to the spectrum collected from '*Ann*'-NDs. This experimental evidence confirmed the effectiveness of the hydrogenation processes and supported the possibility of the role of transfer doping mechanism, ascribable to this surface termination.

The intensity of bands attributed to C-H stretching is larger for '*Ann + Ox + H₂-II*'-NDs than for '*Ann + H₂-I*'-NDs, thus proving that the prior '*Ox*' step guaranteed a higher effectiveness in the hydrogenation process. The lower water-affinity of hydrogenated NDs was highlighted by the absence of relevant features in the 3650 cm$^{-1}$ - 3000 cm$^{-1}$ spectral range and by the parallel drastically decreased intensity of the $\nu_{C=O}$ band. Therefore, the '$H_2$-I' step preserved the hydrophobicity of 'Ann'-NDs, while the '$H_2$-II' step made the 'Ann + Ox'-NDs hydrophobic *via* the elimination of the oxygenated surface groups.

The electrical properties of '$HT$-Ann'-m-NDs were rather peculiar: in opposition to the results reported in Figs. 2 and 3, these latter samples were not influenced at all by water adsorption. The graph of Fig. 4a reports the electrical characterization results for '$HT$-Ann'-NDs, together with those for 'Ann'-NDs for sake of comparison, showing that all of them exhibited a constant impedance value regardless of the relative humidity conditions. The same impedance values were recorded after keeping the NDs in vacuum environment, as it can be seen in Fig. S2a of the SI. It should also be emphasized that '$HT$-Ann'-NDs impedance depended on treatment temperature. The impedance of the '$HT$-Ann-1000$'-NDs varied in the k$\Omega$ range, and it was smaller than the one measured from the '$HT$-Ann-950'-NDs, which varied in the 10 M$\Omega$ range. The latter value was in turn lower than the one measured for the '$HT$-Ann-900'-NDs, which was close to the one measured for the 'Ann'-NDs, *i. e.*, in the G$\Omega$ range.

The electrical characterization results relative to '$HT$-Ann'-NDs can be interpreted by considering the effect of the '$HT$-Ann' process on the surface structure of the nanoparticles. At temperatures greater than 900 °C, structural changes can occur in the NDs even in a chemically inert atmosphere [52]. Therefore, surface graphite can form in '$HT$-Ann'-NDs, contrarily to what happens for 'Ann'-NDs. In this sense, surface graphitization becomes an increasingly prominent feature at rising temperatures, thus explaining the monotonous impedance decrease. The lack of correlation between the impedance values and the relative humidity conditions can be instead interpreted as the consequence of the absence of oxygen-containing moieties at the ND surface, as expected from annealing treatments and indeed confirmed by DRIFT results.

The DRIFT results from '$HT$-Ann'-NDs (shown in Fig. S2b of the SI) evidenced the lack of oxygenated groups, as well as of any other identifiable functionality. Moreover, valuable information could be inferred from the analysis of the Raman results. Fig. 4b reports the Raman spectra acquired from '$HT$-Ann'-NDs together with those from 'Ann'-NDs for sake of comparison. All of the reported spectra were normalized to the intensity of the first-order Raman peak of diamond. It is possible to note that, beside the first-order diamond peak, '$HT$-Ann'-NDs spectra display the G-band at 1580 cm$^{-1}$, which was unequivocally attributed to sp$^2$ carbon hybridization [62], demonstrating that the '$HT$-Ann' treatment indeed caused the formation of graphite at the NDs surface [52]. However, the presence of the first-order diamond peak (observed also in 'Ann'-NDs) indicated that the diamond phase at the core of the nanoparticles was not completely lost upon the '$HT$-Ann' process. The first-order diamond peak was observed at 1317 cm$^{-1}$ in '$HT$-Ann-900'-NDs and '$HT$-Ann-950'-NDs, while it was visible at 1315 cm$^{-1}$ in '$HT$-Ann-1000'-NDs. These values were down-shifted with respect to the peak position measured at 1332 cm$^{-1}$ for the 'Ann'-NDs as well as in bulk diamond [62]. The down-shifted spectral position of the first-order Raman peak of diamond observed in '$HT$-Ann'-NDs was attributed to the structural transformation induced by the high-temperature processing, being probably due to heating of the surface graphitic layer of the nanoparticles induced by laser irradiation, as already reported in a previous work [37]. Finally, it is worth noting that the normalization of the Raman spectra reported in Fig. 4b allows to appreciate that the ratio between diamond and graphitic phases decreased as increasing temperatures of the '$HT$-Ann' treatment. This was indeed indicative of an increasingly pronounced graphitization process, as confirmed from the previously discussed results of the electrical characterization.

### 3.2. Characterization of s-NDs (55 nm median diameter) and μDs (6 μm median diameter)

In order to investigate the effect of diamond particles size on their electrical properties, we performed the processes shown in Fig. 1b also on s-NDs and μDs. The resulting impedance values were displayed in Fig. 5 and summarized in Table 2, together with those resulting from m-NDs (240 nm median diameter), for sake of comparison. Analogously to what observed for m-NDs, the impedance of s-NDs and μDs was not measurable in vacuum. This evidence confirmed that also in this case the presence of surface-adsorbed water represented an essential condition for the onset of electrical conduction. The effectiveness of the above-mentioned processes in the formation of oxygenated groups and hydrogen terminations at the surface of s-NDs and μDs was verified by means of DRIFT spectroscopy, as it can be appreciated from the spectra reported in Fig. S3 of the SI.

The graph reported in Fig. 5a shows the electrical characterization results for the 'Ann + Ox'-samples of different sizes. Electrical conduction appeared to occur *via* the Grotthus mechanism also in s-NDs and μDs since, similarly to what observed from the m-NDs, the impedance values decreased at increasing humidity with a pronounced dependence. From the lower to the upper boundary of the explored RH range, the |Z| value decreased from (64 ± 14) M$\Omega$ to (0.035 ± 0.008) M$\Omega$ in s-NDs, whereas it decreased from (10,000 ± 2000) M$\Omega$ to (42 ± 9) M$\Omega$ for μDs. It is worth noting that in both cases the impedance increasing from high-humidity to low-humidity condition was approximately of three orders of magnitude. It is also worth remarking that the impedance values decreased with decreasing particles size. This could be explained on the basis of the fact that smaller diamond particles are characterized by a larger surface-to-volume ratio: this resulted in a more pronounced water adsorption and thus enhanced conductivity by strengthening Grotthus conduction.

The results of the electrical characterization of the 'Ann + Ox + $H_2$-II'-samples of different sizes were reported in Fig. 5b. As expected, the relatively weak impedance variation as a function of relative humidity that characterized the m-NDs and was interpreted in terms of the transfer-doping conduction was observed also for s-NDs and μDs. In this case, as relative humidity increased, in s-NDs the |Z| value decreased from (5600 ± 1300) M$\Omega$ to (3800 ± 900) M$\Omega$, while in μDs it decreased from (1800 ± 400) M$\Omega$ to (90 ± 20) M$\Omega$. Here the impedance increasing was limited to ~ 1 and ~ 20 factors, respectively. Somewhat surprisingly, the impedance increasing as a function of particle size followed the non-monotonic trend m-NDs → μDs → s-NDs. A possible rationale for this experimental evidence could be formulated by assuming that the contribution from different mechanisms affects electrical conduction. On one hand, holes production is expected to be favoured in smaller diamond particles, due to larger surface-to-volume ratio, resulting into a more pronounced water adsorption that boosts the transfer doping by increasing the electron transfer from the diamond phase to the adsorbed water. On the other hand, it should be taken into consideration that holes are formed subsurface and thus they travel through the diamond material. Holes movement should hence be favoured inside single diamond particles (intraparticle) rather than between different particles (interparticle) and facilitated in large particles, due to the more extended available geometrical path for charge carriers transport. Therefore, the intermediate particle size represents the ideal compromise between the two above-cited mechanisms.

## 4. Conclusions

In the present work, the electrical conduction properties of milled NDs were systematically investigated in relation to both their surface chemistry, size and structural features. We performed electrical characterization, DRIFT and Raman spectroscopy on diamond particles characterized by different median diameter values (*i.e.*, 55 nm, 240 nm and 6 μm), and subjected to annealing, oxidation and hydrogenation thermal treatments.

The electrical characterization of 240 nm NDs revealed that the oxidized NDs were more conductive than the ones treated with annealing at 800 °C. On the other hand, at low relative humidity levels hydrogenated particles displayed an even higher conductivity compared to oxidized ones. Moreover, independently of humidity conditions, the NDs annealed above 900 °C were characterized by a conductivity that was systematically larger than the one observed in the other samples. Thanks to DRIFT spectroscopy, the difference in conductivity between annealed and oxidized NDs was interpreted on the basis of the hydrophilic nature of the latter ones, boosting water adsorption and thus the Grotthuss mechanism. Conversely, the higher conductivity of hydrogenated samples was attributed to the effect of hydrogen terminations, that triggered water-induced transfer doping. In the framework of Raman spectroscopy results, the electrical behavior of high-temperature-annealed NDs was connected to the formation of graphite at the nanoparticles surface. The influence of graphitic phases in the case of the samples subjected to annealing at 800 °C, oxidation and hydrogenation treatments was instead ruled out.

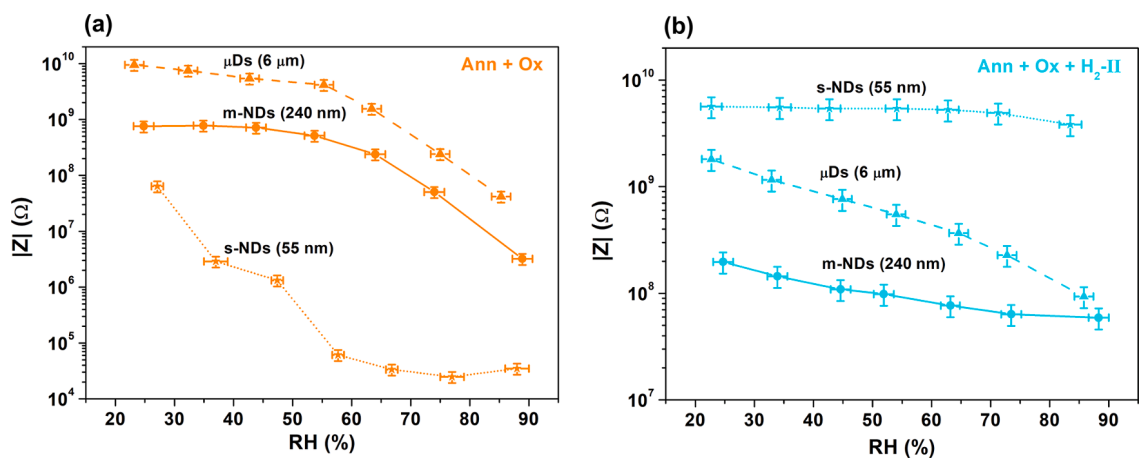

**Fig. 5.** Electrical characterization of s-NDs (55 nm median diameter) and μDs (6 μm median diameter) at different relative humidity (RH) conditions. The data from the m-NDs have been also shown for sake of comparison. (a) '*Ann + Ox.*'-samples. (b) '*Ann +Ox + H₂-II*'-samples.

**Table 2**
Impedance value measured from the s-NDs (55 nm median diameter) and μDs (6 μm median diameter) at relative humidity ~ 25% (indicated as $|Z|_{25}$) and relative humidity ~ 90% (indicated as $|Z|_{90}$). The ratio between $|Z|_{25}$ and $|Z|_{90}$ is also reported. The data from m-NDs (240 nm median diameter) are shown for sake of comparison.

| Sample | $|Z|_{25}$ (MΩ) | $|Z|_{90}$ (MΩ) | $|Z|_{25} / |Z|_{90}$ |
|---|---|---|---|
| *Ann + Ox*-s-NDs | 64 ± 14 | 0.035 ± 0.008 | 1829 |
| *Ann + Ox*-m-NDs | 760 ± 170 | 3.2 ± 0.7 | 238 |
| *Ann + Ox*-μDs | 10,000 ± 2000 | 42 ± 9 | 238 |
| *Ann + Ox + H₂-II*-s-NDs | 5600 ± 1300 | 3800 ± 900 | 1 |
| *Ann + Ox + H₂-II*-m-NDs | 200 ± 40 | 59 ± 13 | 3 |
| *Ann + Ox + H₂-II*-μDs | 1800 ± 400 | 90 ± 20 | 20 |

The results of electrical characterization as a function of particle size showed an increment in conductivity at decreasing NDs size in oxidized samples. This was rationalized as the result of a more pronounced effect of Grotthus mechanism, due to larger surface-to-volume ratio values favouring water adsorption. The higher conductivity of hydrogenated NDs with 240 nm size with respect to both smaller and larger diamond particles was instead explained in terms of a balance between charge carriers production and carriers transport within each particle. However, undoubtedly, our interpretative hypothesis would require a more systematic investigation in order to be confirmed.

Our results demonstrated the connection between heterogeneous NDs features (surface terminations, particle size, structural features) and their electrical properties. They contributed to elucidating the complex combination of the different mechanisms responsible for electrical conduction in NDs. These findings opened the pathway for further studies in the context of the development of nanodiamonds particles for a broad set of different applications. Furthermore, they presented an approach which can be helpful for future research, based on coupling spectroscopic tools with electrical characterization.

## CRediT authorship contribution statement

**Sofia Sturari:** Conceptualization, Investigation, Writing – original draft. **Veronica Varzi:** Investigation. **Pietro Aprà:** Conceptualization, Investigation, Writing – review & editing, Supervision. **Adam Britel:** Resources, Methodology. **Nour-Hanne Amine:** Methodology. **Greta Andrini:** Methodology. **Emilio Corte:** Methodology. **Giulia Tomagra:** Writing – review & editing. **Lorenzo Mino:** Writing – review & editing, Project administration. **Paolo Olivero:** Writing – review & editing, Project administration. **Federico Picollo:** Conceptualization, Supervision, Writing – review & editing, Project administration.

## Declaration of Competing Interest

The authors declare that they have no known competing financial interests or personal relationships that could have appeared to influence the work reported in this paper.

## Data availability

Data will be made available on request.

## Acknowledgments

This work was supported by: "RESOLVE" project funded by the Italian Institute of Nuclear Physics (INFN); "BiophysiX" project funded by the CRT Foundation; "QuantDia" project funded by the Italian Ministry for Instruction, University and Research within the "FISR 2019″ program; "LasIonDef" project funded by the European Research Council under the "Marie Skłodowska-Curie Innovative Training Networks" program; "Departments of Excellence" grant (L. 232/2016) funded by the Italian Ministry of Education, University and Research (MIUR) and 20FUN02 "POLight" project, that has received funding from the EMPIR programme co-financed by the Participating States and from the European Union's Horizon 2020 research and innovation programme.

## Supplementary materials